\documentclass[aps,reprint,superscriptaddress]{revtex4-1}

\usepackage{amssymb}
\usepackage{amsmath}
\usepackage{amscd}
\usepackage{latexsym}

\usepackage{graphicx}
\usepackage{array}

\newcommand{\be}{\begin{equation}}
\newcommand{\ee}{\end{equation}}
\newcommand{\Dlt}{\Delta}
\newcommand{\dlt}{\delta}

\newcommand{\br}{{\bf r}}
\newcommand{\bn}{{\bf n}}
\newcommand{\bfe}{{\bf e}}
\newcommand{\bk}{{\bf k}}
\newcommand{\bh}{{\bf h}}
\newcommand{\bB}{{\bf B}}
\newcommand{\bS}{{\bf S}}
\newcommand{\bt}{\beta}
\newcommand{\vp}{\varphi}
\newcommand{\ep}{\varepsilon}
\newcommand{\al}{\alpha}
\newcommand{\ra}{\rightarrow}

\newcommand{\gm}{\gamma}
\newcommand{\om}{\omega}
\newcommand{\Om}{\Omega}

\newcommand{\rgl}{\rangle}
\newcommand{\lgl}{\langle}

\begin{document}

\title{Regulating spin reversal in dipolar systems by the 
quadratic Zeeman effect} 

\author{V.I. Yukalov}

\affiliation{Bogolubov Laboratory of Theoretical Physics, 
Joint Institute for Nuclear Research, Dubna 141980, Russia}
       
\affiliation{
Instituto de Fisica de S\~{a}o Carlos, Universidade de S\~{a}o Paulo, 
CP 369, S\~{a}o Carlos 13560-970, S\~{a}o Paulo, Brazil}

\author{E.P. Yukalova}

\affiliation{Laboratory of Information Technologies, 
Joint Institute for Nuclear Research, Dubna 141980, Russia}

\begin{abstract}
A mechanism is advanced suggesting the resolution of the dichotomy of long-lived spin 
polarization storage versus fast spin reversal at the required time. A system of 
atoms or molecules is considered interacting through magnetic dipolar forces. The 
constituents are assumed to possess internal structure allowing for the generation 
of the alternating-current quadratic Zeeman effect, whose characteristics can be 
efficiently regulated by quasiresonant dressing. The sample is connected to an 
electric circuit producing a feedback field acting on spins. By switching on and 
off the alternating-current quadratic Zeeman effect it is possible to realize spin 
reversals with a required delay time. The suggested technique of regulated spin 
reversal can be used in quantum information processing and spintronics.       
\end{abstract}

\maketitle

\section{Introduction}

Dipolar interactions are widespread in nature being typical of many biological 
systems \cite{Cameretti_1,Waigh_2}, polymers \cite{Barford_3}, magnetic nanomolecules 
\cite{Kahn_4,Barbara_5,Caneschi_6,Yukalov_7,Yukalov_8,Yukalov_32} and magnetic 
nanoclusters \cite{Kodama_9,Hadjipanayis_10,Yukalov_11,Yukalov_12}. Many dipolar atoms
and molecules can form self-arranged lattices or can be organized in lattice structures 
with the help of external fields 
\cite{Griesmaier_13,Baranov_14,Baranov_15,Gadway_16,Yukalov_17}. Dipolar interactions 
are also typical of ensembles of quantum dots \cite{Escobar_36} and quantum nanowires 
\cite{Corona_37} that possess many properties similar to atoms, because of which
they are often called artificial atoms \cite{Birman_38}. 

Here we consider lattices formed by constituents possessing magnetic dipolar 
moments. These constituents are supposed to enjoy internal structure that can be used
for inducing the alternating-current quadratic Zeeman effect by applying 
quasiresonant linearly polarized light populating internal spin states 
\cite{Cohen_18,Santos_19,Jensen_20,Paz_21}. The alternating-current quadratic 
Zeeman effect can also be induced by quasiresonant linearly polarized microwave 
driving field populating internal hyperfine states \cite{Gerbier_22,Leslie_23,Bookjans_24}. 
It is important that the optically induced quadratic Zeeman effect can also be realized 
with atoms or molecules without hyperfine structure. Such a quasiresonant driving exerts 
quadratic Zeeman shift along the field polarization axis. This shift is described by a
parameter $q_Z$ that does not depend on a stationary external field. By using either 
positive or negative detuning, the sign of the parameter can be varied. The optically or 
microwave induced quadratic Zeeman effect can be easily manipulated and rapidly adjusted, 
thus providing an efficient tool for regulating the properties of the sample.

One of the properties of spin systems, which is extremely important for spintronics, 
as well as for quantum information processing, is the possibility of fast spin reversal. 
At the same time, this property is in contradiction with the other important requirement 
of being able to keep for long time a fixed spin polarization. This is because one 
can fix spin polarization for sufficiently long time by choosing materials with a high 
magnetic anisotropy. However the latter is the major obstacle for realizing fast spin 
reversal. The same dilemma of a well fixed spin polarization versus fast spin reversal, 
which arises in spintronic techniques, also exists in quantum information processing, 
where keeping a fixed spin polarization is necessary for creating memory devices, while 
one needs fast spin reversal for the efficient functioning of such devices. The proposed
devices for realizing quantum computing are also based on spin systems 
\cite{Bernien_39,Zhang_40}.      

Generally, spin reversal in magnetic materials can be induced by inverting a static 
external magnetic field \cite{Shuty_33}. However this is a rather slow process requiring 
sufficiently strong fields. A faster reversal can be realized by applying alternating 
electromagnetic fields, such as produced by lasers \cite{Deb_34,Shelukhin_35}.   

In the present paper, we advance a novel mechanism that, from one side, allows us to 
keep for a long time a fixed spin polarization, while, from the other side, provides 
an efficient tool for realizing a fast spin reversal at any time needed. This mechanism 
suggests a resolution of the dilemma of the fixed spin polarization versus fast spin 
reversal. We show that this can be done for dipolar magnetic systems by employing the 
alternating-current quadratic Zeeman effect.

\section{Scheme of suggested setup}

The suggested setup is as follows. A magnetic sample is inserted into a magnetic coil 
with inductance $L$, containing $n$ turns and having length $l$ and cross-section area 
$A_{coil}$. The coil is a part of an electric circuit also including capacity $C$ and 
resistance $R$. The coil axis is along the axis $x$. A constant external magnetic field 
$B_0$ is directed along the axis $z$. The moving spins of the magnetic sample induce in 
the coil electric current $j$ defined by the Kirchhoff equation
\be
\label{S1}
L \; \frac{dj}{dt} + Rj + \frac{1}{C} \int_0^t j(t')\; dt' = -\; \frac{d\Phi}{dt} \; ,
\ee
in which the electromotive force is caused by the magnetic flux 
$$
\Phi = \frac{4\pi}{c} \; n A_{coil}\; \eta_f \; m_x   
$$
formed by the component of the moving magnetic moment of density $m_x$. Here 
$\eta_f=V/V_{eff}$ is a filling factor being the ratio of the sample volume $V$ to 
the effective volume of the coil $V_{eff}$. The coil inductance is
$$
 L =  4\pi\; \frac{n^2 A_{coil}}{c^2 l} \;  .
$$
The circuit natural frequency, circuit damping, and quality factor are 
\be
\label{S2}
 \om = \frac{1}{\sqrt{LC}} \; , \qquad  \gm = \frac{R}{2L} \; , \qquad 
Q = \frac{\om L}{R} \; .
\ee
The electric current of the coil produces the magnetic field
\be
\label{S3}
H = \frac{4\pi n}{cl}\; j 
\ee
directed along the coil axis. This field, being induced by moving spins, acts back on the spins,
because of which it is called the feedback field. The overall scheme of the suggested setup   
is shown in Fig. 1.

%\Figure 1
\begin{figure}[ht]
\includegraphics[width=6cm]{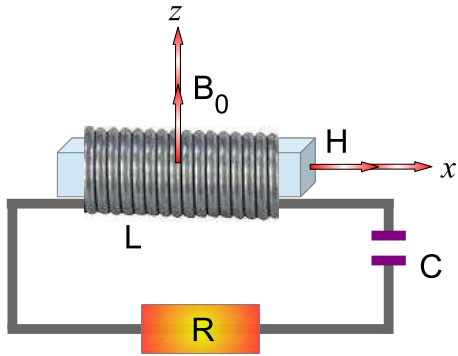}   
\caption{Scheme of suggested setup, as is explained in the text.
}
\label{fig:Fig.1}
\end{figure}

\section{Operator equations of motion}

We consider a system of constituents (atoms or molecules) interacting through dipolar forces. 
The advantage of dealing with such systems is twofold. From one side, as is emphasized in the 
Introduction, systems with dipolar interactions are widespread in nature, hence there exists 
a variety of materials with rather different properties. That is, the system parameters can be 
varied in a wide range. From the other side, dipolar interactions are much weaker then exchange 
interactions, because of which the quadratic Zeeman effect can effectively influence the 
properties of the system. While in hard magnetic materials, such as ferromagnets and 
antiferromagnets, the alternating-current Zeeman effect can be too weak, as compared to the 
energy of exchange interactions, so that the alternating-current Zeeman effect would not 
produce the desired regulation of spin dynamics. 
       
The Hamiltonian of the dipolar lattice system of $N$ sites, each possessing a total 
spin $S$ and characterized by the spin operator ${\bf S}_j$, with $j = 1,2,\ldots, N$, 
is the sum of the Zeeman term $\hat{H}_Z$ and the part $\hat{H}_D$ describing dipolar 
interactions. Generally, dipolar lattices can also include single-site magnetic 
anisotropy. So that the total Hamiltonian is the sum
\be
\label{1}
 \hat H = \hat H_Z + \hat H_D + \hat H_A \;  .
\ee

The Zeeman Hamiltonian contains a linear Zeeman term and a quadratic Zeeman term
induced by the alternating-current quasiresonant light
\cite{Cohen_18,Santos_19,Jensen_20,Paz_21}
\be
\label{2}
  \hat H_Z = - \mu_S \sum_j \bB \cdot \bS_j + q_Z \sum_j ( S_j^z )^2 \; ,
\ee
where $\mu_S = - g_S \mu_B$, with $g_S$ being the spin $g$-factor and $\mu_B$, Bohr 
magneton, while ${\bf B}$ is an external magnetic field acting on spins. The parameter 
$q_Z$ of the quadratic Zeeman effect, induced by a linearly polarized driving field 
coupling internal states, does not depend on the field ${\bf B}$. The axis $z$ is 
assumed to be the polarization axis of the driving field. This parameter $q_Z$, for an
alternating field that is quasiresonant with an internal transition and that is linearly 
polarized along the axis $z$, can be written (see Appendix A) in the form
\be
\label{q}
q_Z = - \frac{\hbar \Omega^2}{4 \Delta_{res}} \; ,
\ee
where $\Omega$ is the driving Rabi frequency and $\Delta_{res}$ is the detuning from an 
internal transition related to spin or hyperfine structure. The parameter $q_Z$ can be 
tailored at high resolution and rapidly adjusted. By applying either positive or negative 
detuning, the sign of this parameter can be made either positive or negative. 

The dipolar Hamiltonian reads as
\be
\label{3}
\hat H_D = 
\frac{1}{2} \sum_{i\neq j} \; \sum_{\al\bt} D_{ij}^{\al\bt} S_i^\al S_j^\bt \; ,
\ee
where the dipolar tensor
\be
\label{4}
 D_{ij}^{\al\bt} = \frac{\mu_S^2}{r_{ij}^3} \; 
\left ( \dlt_{\al\bt} - 3 n_{ij}^\al n_{ij}^\bt \right ) \exp(-\varkappa r_{ij} ) \;  ,
\ee
generally, includes the screening effect, with the screening parameter $\varkappa$. 
The screening of dipolar forces does exist in some materials 
\cite{Jonscher_25,Jonscher_41,Jonscher_42,Tarasov_43,Yukalov_26}, while if it is 
not important, one can set $\varkappa$ to zero. The following consideration does not 
depend on the existence or absence of screening, which is mentioned here only for generality.
Here
$$
r_{ij} \equiv | \; \br_{ij} \; | \; , \qquad  
\bn_{ij} \equiv \frac{\br_{ij}}{r_{ij} } \; , \qquad \br_{ij} = \br_i - \br_j \; .
$$

The total external magnetic field ${\bf B}$ includes a constant field $B_0$ directed 
along the $z$-axis. And the sample is assumed to be placed inside a magnetic coil of an 
electric circuit, so that the coil produces a magnetic feedback field $H$ directed along 
the $x$-axis,
\be
\label{5}
  \bB = B_0\bfe_z + H\bfe_x \; .
\ee

The single-site magnetic anisotropy term can be written \cite{Bedanta_44} in the form
\be
\label{6}
 \hat H_A = - \sum_j D ( S_j^z )^2 \; .
\ee

With the use of the ladder operators $S_j^{\pm} = S_j^x \pm i S_j^y$, the Zeeman term 
transforms into
\be
\label{7}
 \hat H_Z = \sum_j \left [ - \mu_S B_0 S_j^z \; - \; \frac{1}{2} \; \mu_S H
\left ( S_j^+ + S_j^- \right ) + q_Z \left ( S_j^z \right )^2 \right ] \;  .
\ee
And the dipolar part becomes
$$
\hat H_D = \frac{1}{2}  \sum_{i\neq j}  \left [ a_{ij} \left ( S_i^z S_j^z - \;
\frac{1}{2} \; S_i^+ S_j^- \right ) + \right.
$$
\be
\label{8}
\left.
+ b_{ij} S_i^+ S_j^z  +  b^*_{ij} S_i^- S_j^z
+ 2 c_{ij} S_i^+ S_j^z + 2 c^*_{ij} S_i^- S_j^z \right ] \; ,
\ee
in which the interaction coefficients are
$$
a_{ij} \equiv D_{ij}^{zz} \; , \qquad b_{ij} \equiv \frac{1}{4} \left ( D_{ij}^{xx}
- D_{ij}^{yy} - 2i D_{ij}^{xy} \right ) \; ,
$$
\be
\label{9}
 c_{ij} \equiv \frac{1}{2} \left ( D_{ij}^{xz} - i D_{ij}^{yz} \right ) \; .
\ee

Writing down the equations of motion for the spin operators, we introduce the notation
for the Zeeman frequency
\be
\label{10}
 \om_0 \equiv - \frac{\mu_S B_0}{\hbar} > 0 \;  .
\ee
Also we define the quantities
\be
\label{11}  
\xi_i \equiv  \frac{1}{\hbar} \sum_j \left ( a_{ij} S_j^z + c_{ij}^* S_j^-
+ c_{ij} S_j^+ \right )
\ee
and
\be
\label{12}
\vp_i \equiv  \frac{1}{\hbar} \sum_j \left ( \frac{a_{ij}}{2} \; S_j^- -2 b_{ij} S_j^+
-2 c_{ij} S_j^z \right )
\ee
describing local dipolar fields acting on spins. And we introduce the effective force
\be
\label{13}
 f_j \equiv - i \left ( \frac{\mu_S H}{\hbar} + \vp_j \right ) \;  .
\ee

With the above notations, we obtain the spin equations for the transverse spin
$$
\frac{dS_j^-}{dt} = - i ( \om_0 + \xi_j ) S_j^- + f_j S_j^z \; -
$$
\be
\label{14}
 - \; \frac{i}{\hbar} \;
(q_Z - D ) \left ( S_j^- S_j^z + S_j^z S_j^- \right )
\ee
and for the spin $z$-component,
\be
\label{15}
 \frac{dS_j^z}{dt} = - \; 
\frac{1}{2} \left ( f_j^+ S_j^- + S_j^+ f_j \right ) \;  .
\ee

The spin operators in the Heisenberg representation depend on time $t$, which is 
not explicitly shown for the compactness of notations. At the initial moment of time, 
the sample is assumed to be polarized, so that the statistical average of the spin 
$z$-component is nonzero, $\langle S_j^z(0) \rangle \neq 0$.

\section{Dipolar spin waves}

Spin waves are known to exist in ferromagnets and antiferromagnets, where spins interact
through exchange interactions 
\cite{Akhiezer_27,Kalinikos_62,Lvov_63,Gurevich_64,Verba_65,Lisenkov_66}. Here we show that 
spin waves can also exist in the systems with pure dipolar interactions in the presence of 
quadratic Zeeman effect. These spin waves are called dipolar, since they arise in a sample
with purely dipolar interactions, without exchange interactions.

It is necessary to emphasize that the detailed study of spin waves is not our aim here. But 
what is important is to show that they do exist. Their existence is important because it is 
the spin waves that trigger spin motion from a nonequilibrium state. 
  
We keep in mind self-organized spin waves caused by dipolar interactions, but not induced 
by external forces, so that at the initial time, no rotation is imposed on the system,
\be
\label{16}
 \lgl S_j^-(0) \rgl =  \lgl S_j^+(0) \rgl = 0 \;  ,
\ee
and the feedback field has not yet appeared, that is $H = 0$. 

Spin waves are small oscillations around the average spin values, which is described
by representing the spin operators in the form
\be
\label{17}
 S_j^\al = \lgl S_j^\al \rgl + \dlt S_j^\al \;  .
\ee
Due to the property of the dipolar tensor, the interaction functions (\ref{9}) satisfy 
the equality
\be
\label{18}
 \sum_j a_{ij} = \sum_j b_{ij} = \sum_j c_{ij} = 0 \;  .
\ee 
Therefore, for an ideal lattice, where the statistical average does not depend on the 
lattice index, the local fields (\ref{11}) and  (\ref{12}) are actually formed by spin 
waves, since
$$
 \xi_i = \frac{1}{\hbar}  \sum_j \left ( a_{ij} \dlt S_j^z + c_{ij} \dlt S_j^+
+ c_{ij}^* \dlt S_j^- \right ) \; ,
$$
\be
\label{19}
\vp_i = \frac{1}{\hbar}  \sum_j \left ( \frac{a_{ij}}{2} \; \dlt S_j^z - 
2b_{ij} \dlt S_j^+  -2 c_{ij} \dlt S_j^z \right ) \;  .
\ee

Substituting expression (\ref{17}) into the equations of motion, it is necessary to 
be cautious with respect to the last term in Eq. (\ref{14}), taking into account that 
this term is exactly zero for spin $1/2$. Then we use the representation 
\cite{Yukalov_7,Yukalov_8,Yukalov_11,Yukalov_28}
\be
\label{20}
S_j^- S_j^z + S_j^z S_j^- = \left ( 2 - \; \frac{1}{S} \right ) 
\lgl S_j^z \rgl S_j^-
\ee
that is exact for $S = 1/2$ and is asymptotically exact for large spins, when 
$S\ra\infty$.  

Separating in the evolution equations the terms of different orders with respect to 
small spin deviations, in zero order, we have the equations
\be
\label{21}
 \frac{d}{dt} \; \lgl S_j^- \rgl = - i \om_s \lgl S_j^- \rgl \; , \qquad
\frac{d}{dt} \; \lgl S_j^z \rgl = 0 \; ,
\ee
where the effective frequency of spin rotation is
\be
\label{22}
 \om_s \equiv \om_0 + \left ( 2 - \; \frac{1}{S} \right )\; \frac{q_Z-D}{\hbar} \; 
\lgl S_j^z \rgl \;  .
\ee
The first equation gives
$$
 \lgl S_j^-(t) \rgl  = \lgl S_j^-(0) \rgl \; e^{-i\om_s t} \; .
$$
In view of the initial condition (\ref{16}), it follows that
\be
\label{23}
 \lgl S_j^- \rgl = 0 \; , \qquad S_j^- = \dlt S_j^- \; .
\ee
And the second of equations (\ref{21}) shows that $\langle S_j^z \rangle = const$. 

To first order with respect to the spin deviations, we find
\be
\label{24} 
 \frac{d}{dt} \; \dlt S_j^- = - i \om_s \dlt S_j^- - i\vp_j \lgl S_j^z \rgl \; ,
\qquad
 \frac{d}{dt} \; \dlt S_j^z = 0 \; .
\ee
Because of the initial condition $\delta S_j^z(0) = 0$, the above equations give
$\delta S_j^z(t) = 0$. 

Invoking the Fourier transform for the ladder spin operators
$$
S_j^\pm = \sum_k S_k^\pm \exp ( \mp i\bk \cdot \br_j )
$$
and for the interaction functions $a_{ij}$ and $b_{ij}$, 
$$
a_{ij} = \frac{1}{N} \sum_k a_k \exp(i\bk \cdot \br_{ij} ) \; , 
$$
$$
b_{ij} = \frac{1}{N} \sum_k b_k \exp(i\bk \cdot \br_{ij} ) \;   ,  
$$
we reduce the first of equations (\ref{24}) to the form 
\be
\label{25}
 \frac{d}{dt} \; S_k^- =  - i A_k S_k^- + i B_k S_k^+ \; ,
\ee
in which
\be
\label{26}
 A_k \equiv \om_s + \frac{a_k}{2\hbar}\; \lgl S_j^z \rgl \; , \qquad
 B_k \equiv  \frac{2b_k}{\hbar}\; \lgl S_j^z \rgl \; .
\ee
Looking for the solution
\be
\label{27}
 S_k^- = u_k e^{-i\om_k t} + v_k^* e^{i\om_k t} \;  ,
\ee
we obtain the spectrum of spin waves
\be
\label{28}
 \om_k = \sqrt{A_k^2 - |\; B_k\; |^2} \;  .
\ee
Considering the long-wave limit, when $k \ra 0$, we keep in mind that the wavelength
$\lambda = 2 \pi/k$ is much larger than the interspin distance but smaller than the 
sample size. Then the spectrum has the form 
\be
\label{29}
 \om_k \simeq |\; \om_s \; | \left [ 1 - \; \frac{\lgl S_j^z\rgl}{4\hbar\om_s} \;
\sum_{\lgl ij \rgl } a_{ij} (\bk \cdot \br_{ij})^2 \right ] \; .
\ee
Here $<ij>$ implies the summation over the nearest neighbors.

Generally, the spectrum is well defined when $|A_k| > |B_k|$, which yields the 
stability condition
\be
\label{30}
 \left | \; \om_s + \frac{a_k}{2\hbar}\; \lgl S_j^z \rgl \; \right | \; \geq \;  
\left | \;  \frac{2b_k}{\hbar}\; \lgl S_j^z \rgl \; \right | \; .
\ee
Explicitly, this condition reads as
$$
\left | ( 2S - 1) q_Z + \frac{S}{\lgl S_j^z \rgl} \; \hbar \om_0 + \frac{S}{2} \; a_k
- ( 2S - 1 ) D \right | \geq 2 S | b_k| \;   .
$$
This means that spin waves exist when the Zeeman frequency $\omega_0$ and the parameter $q_Z$ 
of the quadratic Zeeman effect are sufficiently large, such that condition (\ref{30}) be valid. 
The quadratic Zeeman effect can stabilize dipolar spin waves \cite{Yukalov_45}. As is clear, 
the existence of dipolar interactions is also crucial.

The occurrence of spin waves is very important, since they serve as a triggering
mechanism initiating spin motion after the system has been prepared in an initial
nonequilibrium state \cite{Yukalov_8,Yukalov_28,Yukalov_29}.

\section{Averaged equations of motion}

Let us consider the temporal behavior of the averaged quantities, the transverse 
spin polarization function
\be
\label{31}
 u \equiv \frac{1}{SN} \sum_{j=1}^N \; \lgl S_j^- \rgl \;  ,
\ee
coherence intensity
\be
\label{32}
  w \equiv \frac{1}{SN(N-1)} \sum_{i\neq j}^N \; \lgl S_i^+ S_j^- \rgl \;  ,
\ee
and the longitudinal spin polarization
\be 
\label{33}
 s \equiv \frac{1}{SN} \sum_{j=1}^N \; \lgl S_j^z \rgl \;  .
\ee

Notice that if one resorts to the standard mean-field approximation, then the averages 
of the local fields (\ref{11}) and (\ref{12}), because of property (\ref{18}), 
become zero, 
$$
 \lgl \xi_j \rgl = \lgl \vp_j \rgl = 0 \;  .
$$
Thus the influence of the dipolar interactions would be lost. However these interactions 
are principally important, since they are necessary for the existence of spin waves 
triggering the initial spin motion. 

To take the dipolar interactions into account, we employ a more refined 
{\it stochastic mean-field approximation} \cite{Yukalov_8,Yukalov_28,Yukalov_30}.
In the process of averaging over the spin variables, we set the notation
$$
\left \lgl \frac{1}{N} \sum_{j=1}^N \xi_j S_j^\al \right \rgl = 
\xi_S \frac{1}{N} \sum_{j=1}^N \; \lgl S_j^\al \rgl \; , 
$$
\be
\label{34}
\left \lgl \frac{1}{N} \sum_{j=1}^N \vp_j S_j^\al \right \rgl = 
\vp_S \frac{1}{N} \sum_{j=1}^N \; \lgl S_j^\al \rgl \; ,
\ee
where $\xi_S$ and $\varphi_S$ are treated as stochastic variables related to local
spin-wave fluctuations. 
   
Realizing statistical averaging over the spin variables, we use the mean-field 
approximation for the spin correlation functions
\be
\label{35}
\lgl S_i^\al S_j^\bt \rgl = \lgl S_i^\al \rgl \lgl S_j^\bt \rgl \qquad
( i \neq j)
\ee
corresponding to spins at different lattice sites. And for the single-site term, we 
employ the decoupling following from Eq. (20),
\be
\label{36}
 \lgl S_j^\al S_j^\bt +  S_j^\bt S_j^\al \rgl =  \left ( 2 - \; \frac{1}{S} \right )
  \lgl S_j^\al \rgl \lgl S_j^\bt \rgl \; ,
\ee
which is exact for $S = 1/2$ and asymptotically exact for $S \ra \infty$. 

The stochastic local fields $\xi_S$ and $\varphi_S$ are defined as random variables 
satisfying the stochastic averaging conditions
$$
\lgl \lgl \xi_S(t) \rgl \rgl = \lgl \lgl \vp_S(t) \rgl \rgl = 0 \; , 
$$
$$
\lgl \lgl \xi_S(t) \xi_S(t') \rgl \rgl = 2\gm_3 \dlt(t-t') \; ,
$$
$$
\lgl \lgl \xi_S(t) \vp_S(t') \rgl \rgl = \lgl \lgl \vp_S(t) \vp_S(t') \rgl \rgl = 0 \; ,
$$
\be
\label{37} 
\lgl \lgl \vp^*_S(t) \vp_S(t') \rgl \rgl = 2\gm_3 \dlt(t-t') \;  ,
\ee
in which $\gamma_3$ is the relaxation rate caused by fluctuating spins interacting 
through dipolar forces. To evaluate the value of $\gamma_3$, we may notice that, in 
view of Eqs. (\ref{37}), the rate $\gamma_3$ can be represented as
\be
\label{38}
 \gm_3 = \left | \; \int_0^\infty \lgl\lgl \xi_S(t) \xi_S(0) \rgl \rgl dt \; \right | \;  .
\ee
The fluctuating field $\xi_S(t)$ behaves according to the law
$$
 \xi_S(t) \propto \gm_2 \exp\{ -i(\om_s -i\gm_2) t\} \;  ,
$$
where $\omega_s$ is the effective spin rotation frequency (\ref{22}) and 
\be
\label{39}
\gm_2 = \frac{1}{\hbar} \; \rho \mu_S^2 S
\ee
is the dipolar transverse attenuation rate, in which $\rho \equiv N/V$ is average spin 
density, with $V$ being the sample volume. The effective spin-rotation frequency (\ref{22}), 
that reads as
\be
\label{40}
  \om_s  =\om_0 + (2S -1) \; \frac{q_Z -D}{\hbar}\; s  \; ,
\ee
can be represented as
\be
\label{41}
 \om_s = \om_0 ( 1 + As) \;  ,
\ee
where the dimensionless parameter
\be
\label{42}
A \equiv (2S - 1) \; \frac{q_Z -D}{\hbar\om_0}
\ee
plays the role of an effective magnetic anisotropy renormalized by quadratic Zeeman effect.

From the integral (\ref{38}), we find
\be
\label{43}
 \gm_3 \cong \frac{\gm_2^2}{\sqrt{\om_s^2 + \gm_2^2} } \;  .
\ee

The effective force (\ref{13}), under averaging over spins, becomes
\be
\label{44}
 f = - i \left ( \frac{\mu_S H}{\hbar} + \vp_S \right ) \;  .
\ee
In the equations of motion, we take into account the existence of the transverse spin 
attenuation rate $\gamma_2$ and the longitudinal attenuation rate $\gamma_1$.

Finally, averaging Eqs. (\ref{14}) and (\ref{15}), we derive the equations for the transverse 
polarization function
\be
\label{45}
 \frac{du}{dt} = - i ( \om_s + \xi_S - i\gm_2 ) u + fs \;  ,
\ee
coherence intensity
\be
\label{46}
 \frac{dw}{dt} = - 2\gm_2 w + ( u^* f + f^* u) s \;   ,
\ee
and the longitudinal spin polarization
\be
\label{47}
 \frac{ds}{dt} = - \; \frac{1}{2} \;  ( u^* f + f^* u) - \gm_1 ( s - s_\infty)  \; ,
\ee
where $s_\infty$ is an equilibrium (or stationary) spin polarization.

\section{Feedback magnetic field}

According to the setup mentioned in Sec. II, the sample is inserted into a coil 
of an electric circuit. Therefore, moving spins induce electric current in the 
coil, which is described by the Kirchhoff equation. In turn, this current creates 
a feedback magnetic field inside the effective coil volume $V_{eff}$. Such a coupling 
with a resonance electric circuit induces in the system the so-called radiation damping 
\cite{Purcell_46,Bloembergen_47,Yukalov_48,Chen_49,Krishnan_50}. The feedback magnetic 
field satisfies the equation \cite{Yukalov_7,Yukalov_8,Yukalov_28,Yukalov_29}
\be
\label{48}
\frac{dH}{dt} + 2\gm H + \om^2 \int_0^t H(t')\; dt' = 
- 4\pi \eta_f \; \frac{dm_x}{dt}
\ee
following form the Kirchhoff equation. Here $\gamma$ is the circuit ringing rate, 
$\omega$ is the circuit natural frequency, and $\eta_f$ is the filling factor 
$\eta_f = V/V_{eff}$. The electromotive force is created by the motion of spins forming 
the magnetic moment with the effective density
$$
m_x = \frac{\mu_S}{V} \sum_{j=1}^N \; \lgl S_j^x \rgl \;   .
$$

Equation (\ref{48}) can be rewritten \cite{Yukalov_7,Yukalov_8,Yukalov_28,Yukalov_29}
as the integral equation 
\be
\label{49}
H = - 4\pi \int_0^t G(t-t') \dot{m}_x(t') \; dt' \;   ,
\ee
in which 
$$
 \dot{m}_x = \frac{N}{2V_{eff}} \; \mu_S S \; \frac{d}{dt} \; (u^* + u ) \; ,
$$
the transfer function is
$$
G(t) = \left [ \cos(\om_{eff}t) - \; 
\frac{\gm}{\om_{eff}} \; \sin(\om_{eff}t) \right ] e^{-\gm t}  \; ,
$$
with the effective frequency
$$
\om_{eff} \equiv \sqrt{\om^2 -\gm^2} \;   .
$$

The electric circuit can be tuned close to the Zeeman frequency $\omega_0$, so that 
the detuning be small,
\be
\label{50}
 \left | \; \frac{\Dlt}{\om} \; \right | \ll 1 \qquad 
( \Dlt \equiv \om - \om_0 ) \;  .
\ee
And, as usual, all attenuations are supposed to be small, such that
\be
\label{51}
 \frac{\gm}{\om} \ll 1 \; , \quad \frac{\gm_1}{\om_0} \ll 1 \; , \quad 
\frac{\gm_2}{\om_0} \ll 1 \; , \quad \frac{\gm_3}{\om_0} \ll 1 \; .
\ee

The coupling between the magnetic coil of the electric circuit and the sample is 
characterized by the coupling rate
\be
\label{52}
 \gm_0 \equiv \frac{\pi}{\hbar}\; \rho \eta_f \mu_S^2 S = \pi \eta_f \gm_2 \;  ,
\ee
which is close to $\gamma_2$, if the volumes of the sample and coil are close to each 
other. Solving Eq. (\ref{49}) by an iterative procedure, to first order with respect to 
the coupling rate, we find
\be
\label{53}
\frac{\mu_S H}{\hbar}  = i ( u X - X^* u^* ) \; ,
\ee
where the coupling function is
$$
X = \gm_0 \om_s \left [ \frac{1-\exp\{-i(\om-\om_s)t-\gm t\}}{\gm+i(\om-\om_s)} \right.
\; +
$$
\be
\label{54}
  + \; \left.
\frac{1-\exp\{-i(\om+\om_s)t-\gm t\}}{\gm-i(\om+\om_s)} \right ] \; .
\ee
When $\omega_s > 0$, the first, quasiresonant, term in the coupling function prevails 
over the second, since
$$
 \left ( \frac{\om - |\om_s|}{\om+|\om_s|} \right )^2 < 1 \;  .
$$
By the same reason, the second term is larger than the first, if $\omega_s < 0$.
Both these cases can be summarized in the expression
$$
X \cong \gm_0 \om_s \; \frac{1-\exp(-i\Dlt_s t - \gm t)}{\gm+i\Dlt_s{\rm sign}\om_s} =
$$
$$
 =
\frac{\gm\gm_0\om_s}{\gm^2+\Dlt^2} \; \left \{
1 - \left ( \cos(\Dlt_s t) -\dlt_s\sin(\Dlt_s t) \right ) e^{-\gm t} \right. \; -
$$
\be
\label{55}
 - \; \left.
i \left [ \dlt_s - \left ( \sin(\Dlt_s t) + \dlt_s\cos(\Dlt_s t) \right ) e^{-\gm t}
\right ] \right \} \;   ,
\ee
where 
$$
 \Dlt_s \equiv \om - |\; \om_s \; | = \om - \om_0 |\; 1 + As \; | \; , 
$$
\be
\label{56}
\dlt_s \; \equiv  \; \frac{\Dlt_s}{\gm} \; {\rm sign}\; \om_s \; .
\ee

\section{Regulating spin reversal}

Substituting the feedback field $H$ into Eq. (\ref{45}) gives the equation
\be
\label{57}
 \frac{du}{dt} = - i \Om u - i \xi_S u - i \vp_S s - X^* u^* s \;  ,
\ee
where
$$
\Om = \om_s - i( \gm_2 - X s) \; .
$$

From Eqs. (\ref{45}) to (\ref{47}) it follows that the functional variable $u$ can 
be classified as fast, while the variables $w$ and $s$ as slow. This allows us to 
employ the scale separation approach \cite{Yukalov_8,Yukalov_28,Yukalov_30} that is 
a variant of the averaging techniques. To this end, we solve 
equation (\ref{57}) for the fast variable $u$ treating there the slow variables $w$ 
and $s$ as quasi-integrals of motion, which yields
$$
u = u_0 \exp \left\{ - i\Om t - i \int_0^t \xi_S(t')\; dt' \right \} -
$$
\be
\label{58}
 - is \int_0^t \vp_S(t') \exp
\left \{ - i \Om(t-t') - i \int_{t'}^t \xi_S(t'')\; dt''\right \} dt' \; .
\ee
The nonresonant counter-rotating term of order $\gamma_2/ \omega_0$ is omitted 
here. Then we substitute the feedback field $H$ and the fast variable $u$ into 
equations (\ref{46}) and (\ref{47}) for the slow variables $w$ and $s$ and average 
these equations over time and over the stochastic variables $\xi_S$ and $\phi_S$. 
This results in the equations for the guiding centers
\be
\label{59}
\frac{dw}{dt} = 2\gm_2 w ( \al s - 1 ) + 2\gm_3 s^2
\ee
and
\be
\label{60}
\frac{ds}{dt} = -\al \gm_2 w  - \gm_3 s - \gm_1 ( s - s_\infty) \; ,
\ee
with the coupling function
$$
 \al \equiv \frac{{\rm Re} X}{\gm_2} = \frac{g\gm^2}{\gm^2+\Dlt_s^2}\; ( 1 + A s )
\times
$$
\be
\label{61}
\times
\left \{ 1 - \left [ \cos(\Dlt_s t) - \dlt_s \sin(\Dlt_s t) \right ] e^{-\gm t}
\right \} \; ,
\ee
in which 
\be
\label{62}
g \equiv \frac{\gm_0\om_0}{\gm\gm_2}
\ee
is the dimensionless coupling parameter characterizing the coupling between the sample
and the electric circuit.  

Analyzing equations (\ref{59}) and (\ref{60}), we take into account that the dipolar 
relaxation rate $\gamma_3$ is smaller then transverse attenuation rate $\gamma_2$, and 
the longitudinal attenuation rate $\gamma_1$ is usually much smaller than $\gamma_2$. 
Measuring time in units of $1/\gamma_2$, we come to the equations
$$
\frac{dw}{dt} = 2w(\al s - 1) + 2 \; \frac{\gm_3}{\gm_2}\; s^2 \; ,
$$
\be
\label{63}
\frac{ds}{dt} = - \al w -\; \frac{\gm_3}{\gm_2} \; s \;  .
\ee
Assume that the system is polarized at the initial time, but no coherence from 
external sources is imposed, so that the initial conditions are
\be
\label{64}
 w(0) = 0 \; , \qquad s(0) = s_0 \;  .
\ee
The external magnetic field $B_0$ at the initial time is directed along the $z$ axis, 
so that the system is in a nonequilibrium state.  
     
The regulation of spin dynamics is based on the possibility of varying in time the 
parameter $q_z$ of the alternating-current quadratic Zeeman effect. The value of 
this parameter can be varied in a rather wide range. For example, dipolar lattices,
organized by means of laser beams 
\cite{Griesmaier_13,Baranov_14,Baranov_15,Gadway_16,Yukalov_17} have the mean 
interatomic distance $a \sim (10^{-5} - 10^{-4})$ cm, hence the average density 
$\rho\sim (10^{12}-10^{15})$ cm$^{-3}$. For the magnetic moments $\mu_S\sim 10\mu_B$, 
the dipolar transverse rate (\ref{38}) is $\gm_2 \sim (10 - 10^{4})$ s$^{-1}$. And 
the value $|q_Z| / \hbar$ can reach $10^5$s$^{-1}$, as can be inferred from Refs. 
\cite{Cohen_18,Santos_19,Jensen_20,Paz_21}.  

There may happen two situations. 

(i) First, if the dipolar system has no single-site 
anisotropy, then one can create a nonzero parameter $q_Z$ for the required time, 
say between zero and $\tau$, during which the initial spin polarization is preserved 
due to the nonzero value of parameter (\ref{42}) that equals
$$
A_0 = (2S - 1) \; \frac{q_Z}{\hbar\om_0} \qquad ( 0 \leq t < \tau ) \;  .
$$
After this, one switches off the quadratic Zeeman effect sending $q_Z$ to zero, 
hence making zero the parameter $A$. This corresponds to the temporal behavior
\begin{eqnarray}
\label{65}
A(t) = \left \{ \begin{array}{ll}
A_0 , ~ & ~ 0 \leq t < \tau \\
0 , ~ & ~ t \geq \tau
\end{array} \right. \; .
\end{eqnarray}

(ii) A similar procedure can be realized when the single-site anisotropy parameter 
$D$ is not zero. Then one can either keep $q_Z$ zero, if the value of $D$ is 
sufficient for freezing the initial spin direction, or create a negative value of 
$q_Z$ for increasing the effective anisotropy to the needed magnitude. After the 
required time $\tau$, one should switch on the quadratic Zeeman effect so that to 
compensate the value of $D$, thus sending $A$ to zero.

%\Figure 2
\begin{figure*}[ht]
\includegraphics[width=15.5cm]{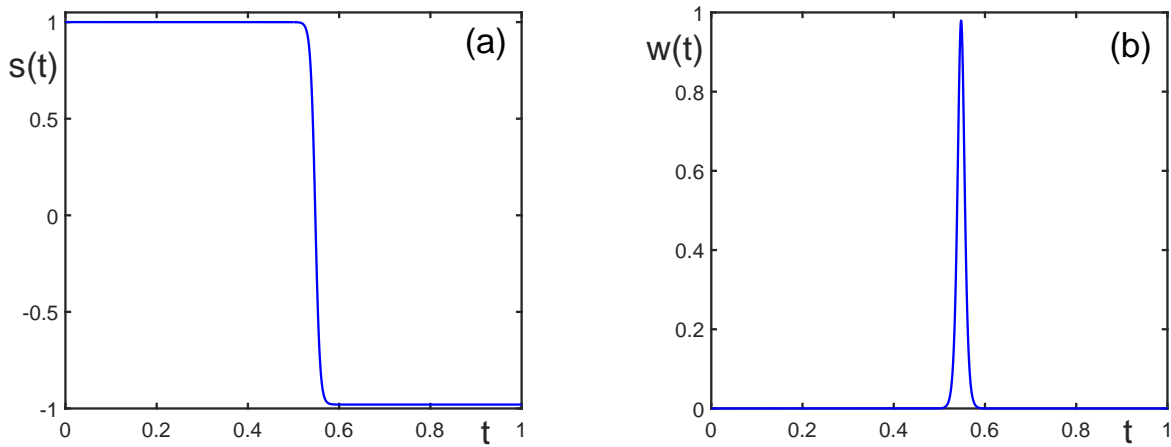}   
\caption{Longitudinal spin polarization (a) and coherence intensity (b) as 
functions of time measured in units of $1/ \gamma_2$. A single spin reversal for 
the parameters $\gm/\gm_2=10$, $\om=\om_0=1000\gm_2$, $A_0=1$,
and $\tau=0.5/\gm_2$.
}
\label{fig:Fig.2}
\end{figure*} 

Numerical solutions to Eqs. (\ref{63}) are presented in Fig. 2, where we set 
$\gm/\gm_2=10$, $\om=\om_0=1000\gm_2$, and $g=100$. For the delay time, we take 
$\tau = 0.5/\gamma_2$, which can be about $10^{-3} - 1$ s. The delay time can be 
taken much longer. As we have checked, under the chosen parameters, the delay time 
$\tau$, during which the spin polarization $s$ practically does not change, can 
reach $100/\gm_2$, which amounts to $0.1-10$ s. The polarization reversal is very 
fast, being approximately equal $\tau_p \approx 1/g s_0 \gamma_2$, which makes 
$10^{-7}-10^{-3}$ s. The polarization reversal is accompanied by a coherent pulse, 
shown in Fig.2b and corresponding to spin superradiance \cite{Yukalov_31}. In that 
way, we achieve the desired goal, being able to keep for long time a fixed 
longitudinal spin polarization, while quickly reversing it as soon as we need.

%\Figure 3
\begin{figure*}[ht]
\includegraphics[width=15cm]{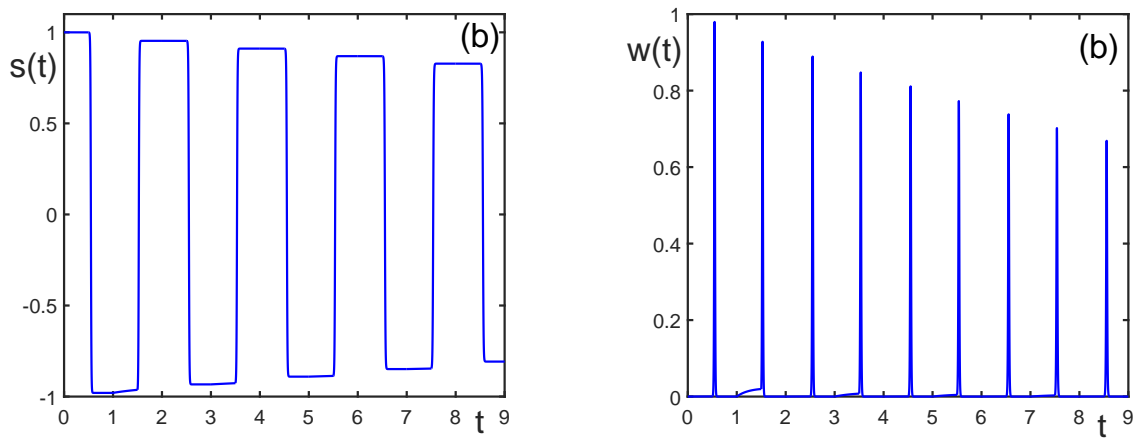}    
\caption{Sequence of longitudinal spin reversals (a) realized by inverting 
the external magnetic field at the moments of the dimensionless time $t_n=(n+1)\tau$, 
measured in units of $1/\gm_2$, and the related sequence of superradiant bursts (b). 
The parameters are as in Fig. 2.
}
\label{fig:Fig.3}
\end{figure*}

Moreover, it is straightforward to repeat the spin reversal several times by 
inverting the external magnetic field $B_0$ during the stage of frozen spin, which 
implies the change of $\om_s$ by $-\om_s$. This procedure, illustrated in Fig. 3, 
goes as follows. The value of $A$ is kept nonzero during the time interval 
$0 \leq t < \tau_1$. At the moment of time $\tau_1$, by regulating the quadratic 
Zeeman effect, the value of $A$ is sent to zero. Thus the first reversal occurs, 
as in Fig. 2. The value of $A$ is kept zero till some time $t_1$. At this time, 
the external field $B_0$ is inverted and $A$ is set to a nonzero value, which is 
kept nonzero till the time $t_1 + \tau_2$. At the moment of time $t_1 + \tau_2$ 
the parameter $A$ is switched off, which results in the second spin reversal. And 
then the process is repeated as many times as necessary. The values $t_n$ and 
$\tau_n$ can be varied, thus realizing the required sequence of spin reversals.

\section{Possibility of experimental implementation}

Choosing appropriate materials for the physical implementation of possible experiments, 
the main point is to select such atoms with internal spin structure that allow for an 
efficient variation by means of the alternating-current Zeeman effect of the dimensionless 
anisotropy parameter $A$ in Eq. (\ref{42}) between small values close to zero and the 
values of order of unity or higher. A collection of such atoms can be arranged in a 
lattice either in a self-organized way or by means of external fields. Also, the atoms 
can be incorporated into a solid-state matrix as a kind of admixture. 

One way is to deal with atomic systems without magnetic anisotropy. For example, one 
can take the atoms of $^{52}$Cr that has the effective spin $S=3$ and magnetic moment 
$6\mu_B$. The nucleus of this atom has zero spin, because of which the atom does not 
possess hyperfine structure, but the alternating Zeeman effect can be induced by a 
quasiresonant light field \cite{Cohen_18, Santos_19}. Since the atomic system does not 
have magnetic anisotropy, the stabilization of an initial nonequilibrium state has to 
be done by the alternating-current Zeeman effect following the procedure explained above 
in paragraph (i). The alternating-current Zeeman parameter $q_Z$ and the Zeeman frequency 
$\omega_0$ should be taken such that the parameter $A$ could reach at least unity.

The other way is to take a system possessing magnetic anisotropy which could be 
compensated for the required time by switching on the alternating-current Zeeman effect 
to provoke the reversal of the magnetization. Consequently, one should follow the way 
described in paragraph (ii). This mechanism sounds more promising for applications in 
view of the smaller energy consumption. 

The solid-state materials, commonly employed in spintronic devices \cite{Geng_62,Li_63}
in the majority of cases correspond to ferromagnetic or antiferromagnetic systems, whose 
spins interact through exchange interactions. If we add to Hamiltonian (\ref{1}) the 
exchange spin term 
$$
 \hat H_{exc} = -\; \frac{1}{2} \sum_{i\neq j} \left [ J_{ij} \left ( S_i^x S_j^x +
 S_i^y S_j^y \right ) + I_{ij} S_i^z S_j^z  \right ] \;  ,
$$
the overall procedure of solving the equations remains the same. The main difference is 
that the effective anisotropy parameter (\ref{42}) now becomes
$$
 A = \frac{1}{\hbar\om_0} \left [ ( 2 S - 1) ( q_Z - D) - S \Dlt J \right ] \;   ,
$$
including the exchange anisotropy 
$$
\Dlt J \equiv \frac{1}{N} \sum_{i\neq j} ( I_{ij} - J_{ij} ) \;   .
$$
In many cases, the latter gives $\Delta J / \hbar \sim 10^{10}$s$^{-1}$. Such a high 
value, to our understanding, cannot be compensated by the alternating-current Zeeman 
effect. 

More promising could be the collections of atoms absorbed on the surface of graphene
\cite{Yaziev_64,Katsnelson_65,Enoki_66,Yukalov_67}. Such adatoms usually also interact 
through exchange forces, but the related magnetic anisotropy can be smaller than in hard 
magnetic materials.      

There exists a large class of magnetic molecules 
\cite{Kahn_4,Barbara_5,Caneschi_6,Yukalov_7,Yukalov_8,Gatteschi_68,Friedman_69,Woodruff_70,
Miller_71,Craig_72,Liddle_73} interacting through dipolar forces, possessing various spins,
between $1/2$ to about $50$, and enjoying a rich internal spin structure. These molecules 
can form ideal self-organized lattices having single-site magnetic anisotropy that can 
stabilize metastable states. 

The lifetime of a magnetic molecule in a metastable state is estimated by the Arrhenius 
law
$$
T_1 = \tau_0 \exp \left ( \frac{U_{eff}}{k_B T} \right ) \; ,
$$
in which $U_{eff} = |D| S^2$ is the effective energy barrier. Clearly, at sufficiently 
low temperatures, lower than a blocking temperature $T_B$, a molecule can be in a 
metastable state for rather long time. For instance, the molecule, labeled as Mn$_{12}$, 
having the spin $S=10$, is characterized by the blocking temperature $T_B = 3 $K, below 
which it has the metastable state lifetime of order $10^7$ s and longer. But the magnetic 
anisotropy of this molecule is too high, with $D/\hbar\sim 10^{11}$ s$^{-1}$. 

Fortunately, there are so many various magnetic molecules that it is possible to find 
among them the molecules with much lower magnetic anisotropy. For example, the molecule, 
labeled as Mn$_{19}$, has the magnetic anisotropy parameter $D/\hbar=7\times 10^6$ 
s$^{-1}$. At the same time, this molecule possesses a very large spin $S=83/2$, so that 
the energy barrier is $U_{eff}/\hbar \sim 10^{10}$ s$^{-1}$. The related blocking 
temperature, for which $U_{eff}$ is much larger than $k_B T$, is $T_B \sim 0.1$ K. 

The effective magnetic anisotropy can be varied by means of mechanical, electric,
and thermal influences \cite{Staunton_74,Heinrich_75}. Also, we can notice that the 
effective magnetic anisotropy parameter $A$ contains the ratio $D/\hbar\omega_0$. 
Therefore the parameter $D$ can be suppressed by increasing the external magnetic field 
$B_0$, that is, by increasing $\omega_0$.   

In order to find out an explicit expression for the reversal time, during which the 
average spin of the system reverses from its initial value $s_0$ to the value about 
$-s_0$, let us consider more in detail the situation, when the effective anisotropy 
parameter $A$ is of the order of one or larger till some time $\tau$, after which this 
parameter $A$ is switched off or suppressed.   

Thus, at the beginning
\be
\label{66}
| \; A \; | \gtrsim 1 \qquad ( t < \tau ) \;  .
\ee
To simplify the following formulas, we take into account the inequalities
$$
 \gm_1 \ll \gm_3 \ll \gm_2 \;  .
$$
Under condition (\ref{66}), we have
$$
\gm_3 = \frac{\gm_2^2}{\om_0|1+As_0|} \; .
$$
The coupling of the sample with the resonant circuit is weak, since
$$
 \al \sim \frac{\gm_0\gm}{\gm_2\om_0|As|} \ll 1  \qquad ( t < \tau) \;  .
$$
We assume that at the initial time no coherent pulses act on the sample, so that $w_0=0$.
Then Eqs. (\ref{63}), with the condition $\gamma_3 t \ll 1$, result in the solution that 
at time $\tau$ gives
\be
\label{67}
 w(\tau) \simeq \frac{\gm_3}{\gm_2}\; s_0^2 \; , \qquad s(\tau) \simeq s_0 \;  .
\ee
 
At time $\tau$ the parameter $A$ is assumed to be suppressed, so that
\be
\label{68}
 |\; A \; | \ll 1 \qquad ( t \geq \tau ) \;  .
\ee
In the case of the resonance, when $\omega =\omega_0$, we have $\omega_s\simeq\omega_0$,
hence $\Delta_s \simeq 0$. For the time $t > \tau$, when $\gamma \tau\gg 1$, the coupling
with the resonator becomes strong, such that $\alpha \simeq g$. The ratio 
$$
 \frac{\gm_3}{\gm_2} = \frac{\gm_2}{\om_0} \ll 1 \qquad ( t \geq \tau) \;  ,
$$
being small, allows us to neglect the term $\gamma_3/\gamma_2$ in Eqs. (\ref{63}). This 
results in the equations
\be
\label{69}
 \frac{dw}{dt} = 2w ( gs - 1) \; , \qquad  \frac{ds}{dt} =  - gw  \; .
\ee
These equations enjoy the exact solution
$$
 w = \left ( \frac{\gm_p}{g\gm_2}\right )^2 
{\rm sech}^2 \left ( \frac{t - t_0}{\tau_p} \right ) \; ,
$$
\be
\label{70}
 s = - \; \frac{\gm_p}{g\gm_2} \; 
\tanh \left ( \frac{t - t_0}{\tau_p} \right ) + \frac{1}{g} \; ,
\ee
in which we return to the time measured in time units. Here $\gamma_p \equiv 1/ \tau_p$ 
and $t_0$ are the integration constants defined by sewing this solution with the values 
(\ref{67}) at the time $t = \tau$. Then, assuming a strong resonator-sample coupling, such 
that $g s_0 \gg 1$, we find
$$
\gm_p =\gm_2 g s_0 \left ( 1 + \frac{\gm_3}{2\gm_2} \right ) \; , 
$$
\be
\label{71}
t_0 = \tau + \frac{\tau_p}{2} \;\ln \; \left (  \frac{4\gm_2}{\gm_3} \right ) \;  .
\ee
The time $\tau_p \equiv 1/\gamma_p$ describes the width of the coherence pulse $w$ and 
also it shows the time during which the spin polarization $s$ reverses form the initial 
value $s_0$ to the final value 
$$
 -\; \frac{\gm_p}{g\gm_2} + \frac{1}{g} \cong -  s_0 \; .
$$
That is, $\tau_p$ is the reversal time, for which we have
\be
\label{72}
 \tau_p = \frac{\gm}{\gm_0\om_0 s_0} \;  .
\ee

In this way, the reversal time depends on the resonator damping $\gamma$ that can be 
varied, the coupling rate $\gamma_0$ that, according to Eq. (\ref{52}), is close to 
$\gamma_2$, the Zeeman frequency $\omega_0$, and the initial spin polarization $s_0$. 
For an external magnetic field $B_0 \sim 1$ T and $\mu_S \sim 10 \mu_B$, we have 
$\omega_0 \sim 10^{11} {\rm s}^{-1}$. Choosing $s_0=1$ and $\gm\sim\gm_0$, we get the 
reversal time $\tau\sim 10^{-11}$ s.

\section{Conclusion}

We have suggested a novel mechanism of regulating spin reversal in a system of atoms 
or molecules possessing internal spin states. The mechanism is based on the use of 
the alternating-current quadratic Zeeman effect occurring when applying quasiresonant
linearly polarized light populating internal spin states. This quasiresonant driving 
exerts quadratic Zeeman shift along the field polarization axis. The optically induced 
quadratic Zeeman effect can be easily manipulated and rapidly adjusted. The appearance 
of the quadratic Zeeman shift is equivalent to the induction of an effective anisotropy 
that can be easily varied. Therefore, it is possible to solve the problem of creating 
a device that could keep spin polarization for long time, but quickly reversing this 
polarization at the required moments of time. The process can be repeated many times, 
producing a sequence of polarization reversals with desired intervals of time.

\section*{Appendix A. Alternating-current Zeeman effect}

The physics of the alternating current quadratic Zeeman effect 
\cite{Cohen_18,Xin_51,Stambulchik_57,Fancher_58,Gan_52} is similar to the alternating 
current Stark effect \cite{Bakos_53,Delone_54,Wu_55,Kien_56}. Let us consider a system 
of atoms enumerated by $j = 1,2,\ldots,N$. Atoms are assumed to be identical, each 
possessing energy levels labeled by an index $n$, with the energies $E_n = \hbar \om_n$ 
and level widths $\gm_n$. In the ground state, a $j$-th atom has the energy $E_a$ and 
spin ${\bf S}_j$. Atoms are subject to an alternating external field that can be written 
as
$$
 \bB_{alt} = \frac{1}{2} \left ( \bh e^{-i\ep t} + \bh^* e^{i\ep t} \right ) \; ,
$$
where $\varepsilon$ is the field frequency. This field interacts with the atomic magnetic 
moment of each atom
$$
 {\bf M}_j = \mu_S \bS_j \;  .
$$
The interaction energy of the field with a $j$-th atom, to first order, is zero on average, 
since the term $-{\bf B}_{alt} \cdot {\bf M}_j$, being averaged over time, is zero. To 
second order of perturbation theory, the interaction energy is 
$$
\Dlt E_j = -\; \frac{1}{4\hbar} \sum_n {\rm Re} \left [ 
\frac{|\lgl \; n\; |\; \bh\cdot{\bf M}_j\; |\; a \rgl |^2}{\om_{na} -\ep -i\gm_{na}} + 
\right.
$$
$$
+ \left.
\frac{|\lgl \; a\; |\; \bh\cdot{\bf M}_j\; |\; n \rgl |^2}{\om_{na} +\ep +i\gm_{na}} 
\right ] \;  ,
$$
with the transition frequencies and transition widths 
$$
\om_{na} \equiv \om_n - \om_a \; , \qquad 
\gm_{na} \equiv \frac{1}{2} \; ( \gm_n + \gm_a ) \;  . 
$$    
The summation goes over all level indices, except $n = a$.  

Let the alternating field be linearly polarized along the axis $z$, so that 
${\bf h} = h_0 {\bf e}_z$. Then, defining the Rabi frequency
$$
 \Om \equiv \frac{|\mu_S h_0|}{\hbar} \;  ,
$$
we have 
$$
\Dlt E_j = -\; \frac{\hbar\Om^2}{4} \sum_n |\lgl a \; | \; S_j^z\; | \; n \rgl |^2
\times
$$
$$
\times
 \left [ \frac{\om_{na}-\ep}{(\om_{na}-\ep)^2+\gm_{na}^2} + 
 \frac{\om_{na}+\ep}{(\om_{na}+\ep)^2+\gm_{na}^2} \right ] \; .
$$

The alternating field is tuned close to one of the transition frequencies, corresponding 
to some fixed $n = b$, so that the quasiresonance condition be valid
$$
 \left | \; \frac{\Dlt_{res}}{\om_{ba}} \; \right | \ll 1 \qquad 
( \Dlt_{res} \equiv \om_{ba} -\ep ) \;  .
$$

Then, taking into account the identity
$$
 \sum_n \; |\lgl a\; | \; S_j^z \; | \; n \rgl |^2 = 
\sum_n \; \lgl a\; |\; S_j^z \; | \; n \rgl  \lgl n \; |\; S_j^z \; | \; a \rgl =
$$
$$
=
\lgl a\; |\; ( S_j^z )^2 \; | \; a \rgl \; ,
$$
we come to the expression
$$
\Dlt E_j \cong -\; \frac{\hbar\Om^2\Dlt_{res}}{4(\Dlt_{res}^2+\gm_{ba}^2)} \;
\lgl a\; |\; ( S_j^z )^2 \; | \; a \rgl \; .
$$

The Hamiltonian of the effect for a $j$-th atom is defined as the operator whose 
quantum-mechanical average yields the additional energy 
$$
 \Dlt E_j \equiv  \lgl a\; |\; \Dlt \hat H_j \; | \; a \rgl \; .
$$
This results in the Hamiltonian
$$
 \Dlt \hat H_j = - \; \frac{\hbar\Om^2\Dlt_{res}}{4(\Dlt_{res}^2+\gm_{ba}^2)} \;
(  S_j^z )^2 \; .
$$
Respectively, the corresponding Hamiltonian term for the whole collection of $N$ atoms 
is
$$
 \hat H_{QZ} = \sum_j \Dlt \hat H_j \;  .
$$

As is evident, it would not be reasonable to take the exact resonance condition 
$\Dlt_{res}=0$, since then the interaction energy tends to zero. Therefore on takes 
$\varepsilon$ not too close to the transition frequency, in the sense that the 
off-resonance condition be true,
$$
 \left | \; \frac{\Dlt_{res}}{\gm_{ba}} \; \right | \gg 1 \; .
$$
Under this condition, the Hamiltonian term becomes
$$
 \Dlt \hat H_j = - \; \frac{\hbar\Om^2}{4\Dlt_{res}} \; (  S_j^z )^2 \; .
$$

Finally, summing over all atoms in the system, we get the interaction term corresponding
to the alternating-current quadratic Zeeman effect
$$
 \hat H_{QZ} = q_Z  \sum_j (  S_j^z )^2 \; ,
$$
with the parameter $q_Z$ defined in Eq. (\ref{q}).

In order to exhibit the alternating-current Zeeman effect, an atom, or molecule, 
needs to possess an internal spin structure. If the nucleus of an atom has a nonzero 
spin, then there exists hyperfine structure. And even if there is no the latter, when 
the nuclear spin is zero, there always exists the spin structure of energy levels, as 
soon as an atom contains electrons
\cite{Cohen_18,Santos_19,Jensen_20,Paz_21,Gerbier_22,Leslie_23,Bookjans_24,Xin_51,
Stambulchik_57,Fancher_58,Gan_52}. Since all atoms have electrons, their energy levels 
depend on the presence of external magnetic fields, including alternating fields. 
Therefore, the alternating-current Zeeman effect occurs for practically all atoms and 
molecules \cite{Grigoriev_59,Lajunen_60,Aller_61}.

\end{document}